\documentclass[%
reprint,
 amsmath,amssymb,
 aps,
 pra,
]{revtex4-1}

\usepackage{graphicx}
\usepackage{epsfig}
\usepackage{epstopdf}
\usepackage{subfigure,color}
\usepackage{dcolumn}
\usepackage{bm}
\usepackage{soul}

\def\_#1{{\bf #1}}
\def\@#1{_{\rm #1}}
\def\Re{{\rm Re\mit}}
\def\Im{{\rm Im\mit}}

\begin{document}

\title{Dipole polarizability of time-varying particles}

\author{M.~S.~Mirmoosa$^{1,2}$}
\email{mohammad.mirmoosa@aalto.fi}
\author{T.~T.~Koutserimpas$^{1}$}
\author{G.~A.~Ptitcyn$^{2}$}
\author{S.~A.~Tretyakov$^{2}$} 
\email{These authors jointly supervised this work.}
\author{R.~Fleury$^{1}$}
\email{These authors jointly supervised this work.}

\affiliation{$^{1}$Laboratory of Wave Engineering, Swiss Federal Institute of Technology in Lausanne (EPFL), CH-1015 Lausanne, Switzerland\\$^{2}$Department of Electronics and Nanoengineering, Aalto University, P.O.~Box 15500, FI-00076 Aalto, Finland}

\begin{abstract}
Invariance under time translation (or stationarity) is probably one of the most important assumptions made when investigating electromagnetic phenomena. Breaking this assumption is expected to open up novel possibilities and result in exceeding conventional limitations. However, to explore the field of time-varying electromagnetic structures, we primarily need to contemplate the fundamental principles and concepts from a nonstationarity perspective. Here, we revisit one of those key concepts: The polarizability of a small particle, assuming that its properties vary in time. We describe the creation of induced dipole moment by external fields in a nonstationary, causal way, and introduce a complex-valued function, called temporal complex polarizability, for elucidating a nonstationary Hertzian dipole under time-harmonic illumination. This approach can be extended to any subwavelength particle exhibiting electric response. In addition, we also study the classical model of the polarizability of an oscillating electron using the equation of motion whose damping coefficient and natural frequency are changing in time. Next, we theoretically derive the effective permittivity corresponding to time-varying media (comprising free or bound electrons, or dipolar meta-atoms) and explicitly show the differences with the conventional macroscopic Drude-Lorentz model. This paper will hopefully pave the road towards better understanding of nonstationary scattering from small particles and homogenization of time-varying materials, metamaterials, and metasurfaces.     
\end{abstract}

\maketitle


\section{Introduction}

Temporal modulation~\cite{Faraday} in electromagnetic systems (e.g.,~Refs.~\cite{Cullen,Oliner,Tien,Morgenthaler,Kamal,Currie,Edmondson,Anderson,Simon,Holberg}) is an efficient technique to achieve exotic wave phenomena and intriguing functionalities. Nonreciprocity and isolation~\cite{TM-nonreciprocity1,TM-nonreciprocity2,TM-nonreciprocity3,TM-nonreciprocity4,TM-isolation1,TM-isolation2,XuchennnnWWW}, frequency conversion and generation of higher-order frequency harmonics~\cite{TM-frequency1,TM-frequencytranslationGrbic,TM-frequencyconvMosallaeiSalary}, wavefront engineering~\cite{TM-frequency1,TM-wavefront1,TM-wavefront2}, one-way beam splitting~\cite{TM-beam}, extreme accumulation of energy~\cite{TM-Energy}, parametric amplification~\cite{Cullen,Tien,Fleury222,Fleury333}, and wideband impedance matching~\cite{TM-Matching} are some of those functionalities that have been reported in the past. One possibility that time modulation can provide is to instantaneously control the radiation from subwavelength particles~\cite{TM-Radiation1,TM-Radiation2}. This is due to the fact that  electric and magnetic dipole moments, $\_p(t)$ and $\_m(t)$,  induced in a particle under illumination, can be temporally engineered in a desired fashion. In general, both  the geometry of the particle and the optical properties of the material from which the particle is made can be  properly modulated in time by an external force. 

From the stationary perspective, it is assumed that the particle is static, and its characteristics do not vary in time. As a consequence, the induced dipole moments are conventionally described in the frequency domain simply through the complex dyadic electric and magnetic polarizabilities~\cite{polarizability-tensors}: 
\begin{equation*}
\begin{split}
&\overline{\_p}=\overline{\overline{\alpha}}_{\rm{ee}}(\omega)\cdot\overline{\_E}+\overline{\overline{\alpha}}_{\rm{em}}(\omega)\cdot\overline{\_H},\cr
&\overline{\_m}=\overline{\overline{\alpha}}_{\rm{me}}(\omega)\cdot\overline{\_E}+\overline{\overline{\alpha}}_{\rm{mm}}(\omega)\cdot\overline{\_H}.
\end{split}
\end{equation*}
Here, $\overline{\_E}$ and $\overline{\_H}$ are the Fourier transforms of the external electric and magnetic fields, respectively. However, the above equations cannot be generally applied for a dynamic particle, because the very definition of frequency-domain parameters is based on the assumption that the particle is stationary. In fact, concerning a time-varying particle, we need to return to the time domain, and, subsequently, revisit the description of the instantaneously induced dipole moments in terms of the dyadic polarizabilities model. An appropriate  description should explicitly indicate the nonstationary characteristic of the problem, along with the linearity and memory (frequency dispersion). The importance of this  study is not limited only to the understanding and engineering of instantaneous radiation, but it is also important for the proper characterization and realistic implementation of time-varying metamaterials and  metasurfaces~\cite{CalozTVMM1,CalozTVMM2}, because they are formed by time-varying meta-atoms. Therefore, having a clear picture about the polarizability of meta-atoms paves the road towards homogenization models~\cite{SihvolaMF,Enghetanano} taking into account nonstationarity, and its interplay with dispersion phenomena. 

In this paper, we thoroughly scrutinize the concept of polarizability associated with a particle which is varying in time. For simplicity, we assume that the particle has only isotropic electric response. We analytically study how to determine the polarizability of such a particle when it is located in free space. For this study, we employ the Hertzian dipole model which is a conventional model to describe a stationary resonant particle with electric response. Determining the polarizability, we also explain the interaction of  nonstationary dipoles with waves in terms of the particle polarizability. Furthermore, we move forward and consider the particle as a constituent of a time-varying material. Regarding this scenario, we focus on the classical bound electron (as the particle under study) and derive the corresponding polarizability by assuming a time-dependent damping coefficient and natural frequency in the equation of motion. Accordingly, we obtain the nonstationary Drude-Lorentz model for an effective medium and show how fundamentally different this new model is from the conventional model written for a stationary medium.

The paper -- as a foundational step towards understanding of nonstationary scattering from small particles and  time-varying (artificial) media -- is organized as follows: In Section~\ref{sec:basicconcepts}, we give a fundamental description of polarization of an arbitrary time-varying dipolar particle as a response to the excitation field by using the concept of electric polarizability. Since this description (initially inspired by the analysis and synthesis of linear time-varying systems in communications and control engineering~\cite{communicationengLTVS,ZADEHAIP}) is not well covered in the  literature and is missing in the classical electrodynamics textbooks~\cite{Landauref,TEXTBOOKSLTIS,jackson},  it helps the reader to obtain a proper perspective and is essential for understanding of the other sections of the paper. In the same section, we additionally discuss causal models of effective material parameters of time-varying media, following the same principles as for a single time-varying particle. Next, in Sections~\ref{Appendix:nsdpktcp} and~\ref{electron}, under nonstationary conditions, we treat small particles and classical electrons based on their corresponding polarizabilities. Section~\ref{electron} is also devoted to effective material models of nonstationary media. Finally, Section~\ref{secconclusion} concludes the paper. 


\section{Basic Concepts}
\label{sec:basicconcepts}

\subsection{Polarizability kernel}

For a linear and stationary subwavelength particle with electric response, there is a temporally nonlocal connection between the instantaneous electric dipole moment $\mathbf{p}(t)$ and the exciting electric field $\mathbf{E}(t)$. This connection is described by a convolution integral as
\begin{equation}
\_p(t)=\int_0^\infty\alpha(\gamma)\_E(t-\gamma)d\gamma,
\label{eq:introduction}
\end{equation}
where $\alpha(\gamma)$ is a time-dependent function called electric polarizability kernel (here, we assume that the dipole and the field are parallel and there is no bianisotropy). The above equation illustrates two notable characteristics. The first is that if the electric field is temporally shifted by $t_{\rm{sh}}$, the dipole moment will be also shifted by the same time $t_{\rm{sh}}$ due to the stationarity of the particle. In other words, 
\begin{equation}
\_p(t-t_{\rm{sh}})=\int_{0}^\infty\alpha(\gamma)\_E(t-\gamma-t_{\rm{sh}})d\gamma.
\end{equation}
The second characteristic, associated with causality, states that the instantaneous dipole moment at a certain time depends on the field at that time and the
evolutionary progress over  past times. 

The situation is very different if the particle under study is changing in time. Causality is certainly a fundamental concept in nature which should be scrutinised carefully. However, the first characteristic, having to do with invariance with respect to translations in time, is not true anymore. For interactions of nonstationary  particles with fields, a temporal shift of the electric field does not result in an equivalent temporal shift of the induced dipole moment. We should use a more general linear and causal relation between the induced dipole moment and the exciting field, which we write as 
\begin{equation}
\_p(t)=\int_{0}^{+\infty}\alpha(\gamma,t)\_E(t-\gamma)d\gamma.
\label{eq:polar}
\end{equation}
Here, in contrast to Eq.~\eqref{eq:introduction}, the polarizability kernel $\alpha$ is not only a function of the delay time between the action and reaction ($\gamma$), but it also depends on the observation time ($t$). In other words, this formula means that at every moment of time $t$, we deal with a different particle and with a different frequency dispersion rule (defined by the integral kernel as a function of $\gamma$). As a consequence of that, the instantaneous value of the dipole moment depends not only on the past and present values of the exciting field, but also on the whole history of evolution of the particle properties.

Based on Eq.~\eqref{eq:polar}, let us discuss the physical  meaning of the polarizability of a nonstationary particle. If the electric field is chosen to be the Dirac delta function $\_E(t)=\delta(t-t_0)\_u$ ($\_u$ is a unit vector), the dipole moment equals
\begin{equation}
\_p(t)=\alpha(t-t_0,t)\_u.
\end{equation}
In other words, the polarizability $\alpha$ is the impulse response of the dipole. As we see, in the nonstationary situation, the impulse response depends, as in usual stationary linear systems, on how much time has passed since the pulse excitation was applied, but also on time explicitly. This property clearly manifests the fact that the particle responds differently at different moments of time. 

Beside using Eq.~\eqref{eq:polar}, it is sometimes convenient to apply an alternative integral form to describe the dipole moment (for example, see Sections~\ref{Appendix:nsdpktcp} and \ref{electron}). Let us consider the following independent variable: $\tau=t-\gamma$. By changing variable in Eq.~\eqref{eq:polar} and defining 
\begin{equation}
h(t,\tau)=\alpha(\gamma,t)\vert_{\gamma=t-\tau},
\end{equation}
we can equivalently write 
\begin{equation}
\_p(t)=\int_{-\infty}^{t}h(t,\tau)\_E(\tau)d\tau.
\label{causaleq:dp}
\end{equation}
In this alternative representation of causal linear relations, the $\tau $ variable has the meaning of time moments in the past, and the chosen integration limits ensure that the induced dipole does not depend on the field values in the future. Also, in this form, assuming delta-function excitation $\_E(\tau)=\delta(\tau-t_0)\_u$, we find the impulse response in general form 
$\_p(t)=h(t,t_0)\_u$.
Notice that in the stationary scenario, the function $h(t,\tau)$ dependents only on the time difference between the observation time and a time moment in the past: $h(t,\tau)=h(t-\tau)$. Consequently, the integral expressed by Eq.~\eqref{causaleq:dp} becomes simply a convolution, and the dependency of the polarizability kernel on the observation time $t$ vanishes, i.e.,~the polarizability kernel is only written in terms of the variable $\gamma$: $\alpha(\gamma,t)=h(t,t-\gamma)=h(\gamma)$. Accordingly, we obtain Eq.~\eqref{eq:introduction} which was explained in the beginning of this section.

At this point, where the functions $\alpha(\gamma,t)$ and $h(t,\tau)$ and the corresponding relation between them have been  discussed, it is better to build a general consensus that we refer only to the function $\alpha(\gamma,t)$ as the polarizability kernel. This agreement will help us to avoid any confusion throughout the paper. Therefore, when working with Eq.~\eqref{causaleq:dp} and the function $h(t,\tau)$, remember that we need to do a simple algebraic manipulation and determine $\alpha(\gamma,t)$ in order to present the polarizability kernel.

\subsection{Temporal complex polarizability}

\subsubsection{Definition}

\begin{figure}
\centering
\includegraphics[width=1\linewidth]{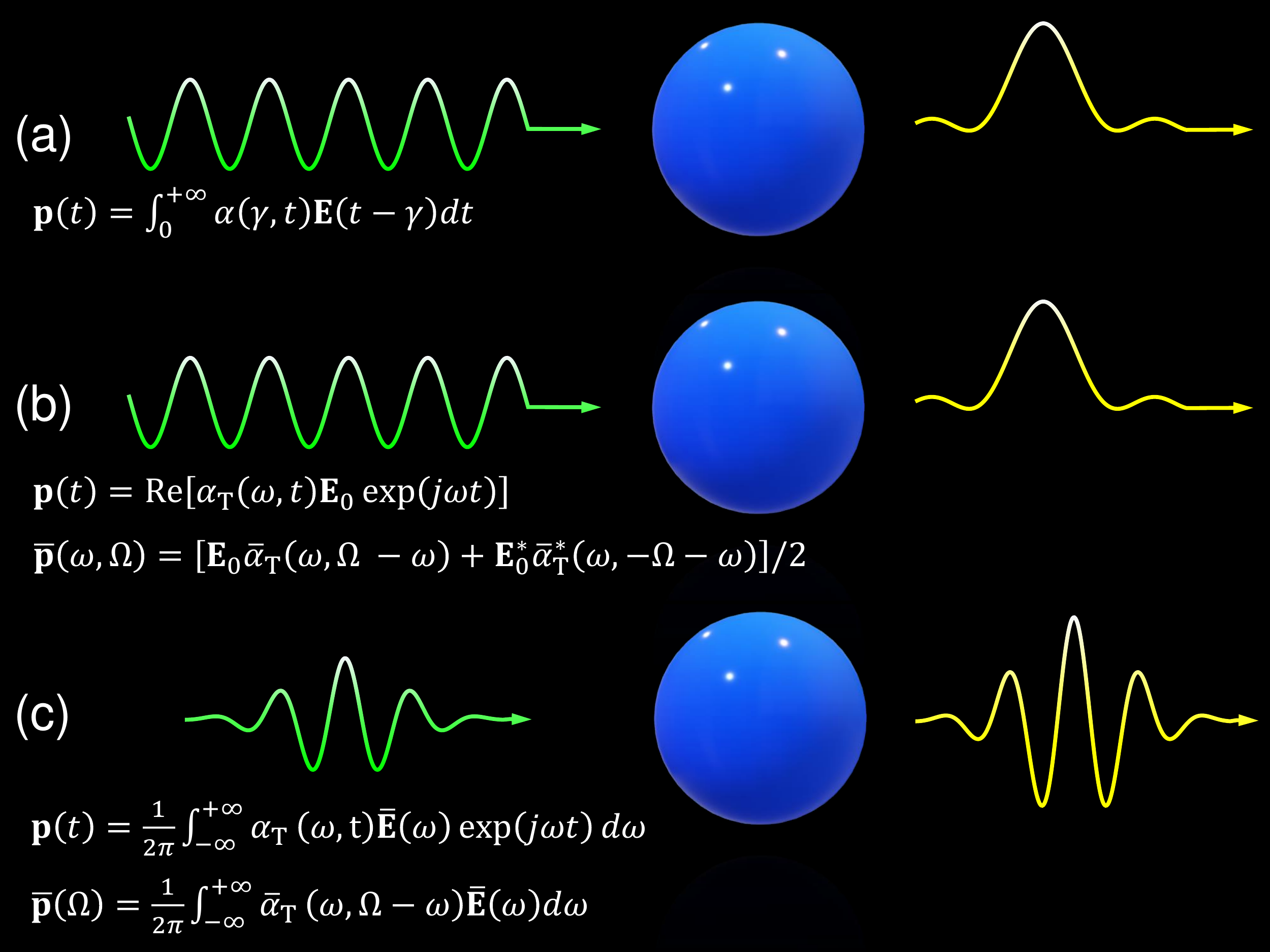}
\caption{Schematic view of wave scattering by a time-varying spherical particle. We assume that, for example, the optical properties of the particle is changing in time. (a)--The particle under time-harmonic illumination. We use the general concept of polarizability kernel $\alpha(\gamma,t)$ in order to study this scattering problem. (b)--The particle under the same time-harmonic illumination. However, we employ the particular concept of temporal complex polarizability $\alpha_{\rm{T}}(\omega,t)$ and its Fourier transform $\overline{\alpha}_{\rm{T}}(\omega,\Omega)$ for studying the corresponding problem. (c)--The particle under arbitrary signal illumination. Here, temporal complex polarizability and its Fourier transform are still valid to be used.} 
\label{fig1schmlsnp}
\end{figure}

Let us consider a time-harmonic excitation by a given electric field $\_E(t)={\rm{Re}}\Big[\_E_0\exp(j\omega t)\Big]$. Here, $\_E_0$ denotes the complex amplitude, $\omega$ is the angular frequency, and  $\Re$ means the real part of the expression inside the brackets. The reason for choosing the time-harmonic excitation is the fact that we want to concentrate on understanding the effects of time variations of the particle itself, and it is convenient to use the simplest possible exciting fields. Since the particle is linear, response to arbitrary excitation can be found using the Fourier expansion of the incident field.  Therefore, it is logical to create a model for time-harmonic excitation. 

Since from the beginning we have assumed that the field and the dipole moment are parallel, the polarizability kernel is a scalar value. Substituting the electric field into Eq.~\eqref{eq:polar}, we find that 
\begin{equation}
\_p(t)={\rm{Re}}\Big[\alpha_{\rm{T}}(\omega,t)\cdot \_E_0\exp(j\omega t)\Big],
\label{eq:dmpolth}
\end{equation}
where
\begin{equation}
\alpha_{\rm{T}}(\omega,t)=\int_{0}^{+\infty}\alpha(\gamma,t)\exp(-j\omega\gamma)d\gamma. 
\label{mostimeq}
\end{equation}
The instantaneous electric dipole moment is the real part of a {\it complex-valued function} which is multiplied by the complex amplitude of the time-harmonic electric field $E_0\exp(j\omega t)$. This is in clear analogy with the conventional stationary case in which the instantaneous dipole moment is the real part of the complex-valued, frequency-domain electric polarizability multiplied by the electric field amplitude and the time-harmonic exponential factor. However, here, the complex function $\alpha_{\rm{T}}$ depends on the time variable $t$. Thus, we name such function as ``temporal complex polarizability". The index ``T", reminding ``temporal", distinguishes the function $\alpha_{\rm{T}}$ from the polarizability kernel $\alpha$. 

Recall that the above definition is for the case of time-harmonic excitation (which is also indicated in  Figure~\ref{fig1schmlsnp}). As one can realize in accordance with Eq.~\eqref{eq:dmpolth}, by using this definition, the complexity of the problem dramatically reduces, and the induced dipole moment is simply described based on the temporal complex polarizability $\alpha_{\rm{T}}(\omega,t)$. The result of such simplicity is clear, for example, in the next section -- Section~\ref{secHertziandipole} -- where we discuss the interaction of the point electric dipole with time-harmonic incident electric fields. We will see how the instantaneous power exerted on the dipole and the instantaneous power scattered from the dipole are elegantly written in terms of the temporal complex polarizability (without making any integration). Furthermore, the importance and advantage of the aforementioned simplicity can be also understood in more complicated problems such as homogenization of time-varying artificial media (metamaterials), in which the effective macroscopic parameters should be derived, and, subsequently, the corresponding dispersion relations need to be calculated. Regarding those problems, therefore, it is quite reasonable to employ the concept of the temporal complex polarizability $\alpha_{\rm{T}}(\omega,t)$ instead of the polarizability kernel $\alpha(\gamma,t)$ in the time domain. We emphasize that due to the linearity, the temporal complex polarizability can be used for nonharmonic excitations as well by applying the Fourier expansion (see Figure~\ref{fig1schmlsnp}). 

\subsubsection{Properties} 

Let us describe some features which are inferred from Eqs.~\eqref{eq:dmpolth} and \eqref{mostimeq}. By contemplating Eq.~\eqref{mostimeq}, we firstly see that the temporal complex polarizability is the Fourier transform of the polarizability kernel with respect to the temporal variable $\gamma$. Secondly, because the functions $\_p(t)$, $\_E(t)$, and $\alpha(\gamma,t)$ are real-valued, based on Eq.~\eqref{mostimeq} and for real angular frequencies, we deduce that 
\begin{equation}
\alpha_{\rm{T}}^\ast(\omega,t)=\alpha_{\rm{T}}(-\omega,t), 
\end{equation}
in which $^\ast$ represents the complex conjugate. Furthermore, similar to the stationary scenario, in Eq.~\eqref{mostimeq}, the integration is over the positive half-axis of $\gamma$, which reflects causality of the system and indicates that the temporal complex polarizability obeys Kramers-Kronig relations~\cite{Landauref}.

About Eq.~\eqref{eq:dmpolth}, it explicitly confirms the expectations that the dipole moment induced by time-harmonic fields is not necessarily time-harmonic. In Ref.~\cite{TM-Radiation1}, the authors have recently discussed this fact without studying the polarizability. Importantly, the temporal variations of the  dipole moment can be in principle fully engineered (while the excitation field is time-harmonic) only by choosing the proper temporal variation of the particle modulation. Beside this property, it is instructive to take the Fourier transform of  Eq.~\eqref{eq:dmpolth}. For that, this equation can be rewritten as  
\begin{equation}
\_p(t)={1\over2}\Bigg[\_E_0\alpha_{\rm{T}}(\omega,t)\exp(j\omega t)+\_E_0^\ast\alpha_{\rm{T}}^\ast(\omega,t)\exp(-j\omega t)\Bigg].
\label{eq:twOmega}
\end{equation}
Now, by defining the usual Fourier transform of an arbitrary temporal function $g(t)$ as $\overline{g}(\Omega)=\int_{-\infty}^\infty g(t)\exp(-j\Omega t)dt$ and applying this operation to Eq.~\eqref{eq:twOmega}, we obtain 
\begin{equation}
\overline{\_p}(\omega,\Omega)={\_E_0\overline{\alpha}_{\rm{T}}(\omega,\Omega-\omega)+\_E_0^\ast\overline{\alpha}_{\rm{T}}^\ast(\omega,-\Omega-\omega)\over2}. 
\label{eq:2angfrerev}
\end{equation} 
The first argument of the Fourier transform of $\_p(t)$ indicates that this expression is derived for excitation by a time-harmonic electric field at frequency $\omega$. For excitations at other frequencies, the time dependence of the induced dipole moment $\_p(t)$ will be different. The dependence on two frequency variables is intriguing:  The first frequency argument, $\omega$, corresponds to the Fourier transform with respect to the variable $\gamma$, and the second angular frequency, $\Omega$, is due to the Fourier transform with respect to the variable $t$. Notice that since the dipole moment $\_p(t)$ is real-valued, we have  $\overline{\_p}^\ast(\omega,\Omega)=\overline{\_p}(\omega,-\Omega)$. This relation can be readily proved by using Eq.~\eqref{eq:2angfrerev}. It is worth noting that $\alpha_{\rm{T}}(\omega,t)$ does not obey this relation, because $\alpha_{\rm{T}}(\omega,t)$ is not necessarily a real-valued function. Therefore, in general, 
\begin{equation}
\overline{\alpha}^\ast_{\rm{T}}(\omega,\Omega)\neq\overline{\alpha}_{\rm{T}}(\omega,-\Omega). 
\end{equation}


\subsubsection{Temporal complex susceptibility and permittivity} 

Before moving to the next section, it is worth noting that in analogy with the dipole moments, a similar time-domain description should be used for the electric and magnetic flux densities $\_D(t)$ and $\_B(t)$ as linear and causal functions of $\_E(t)$ and $\_H(t)$. If a medium is stationary, we readily write~\cite{polarizability-tensors} 
\begin{equation}
\begin{split}
&\overline{\_D}=\overline{\overline{\epsilon}}(\omega)\cdot\overline{\_E}+\overline{\overline{\xi}}(\omega)
\cdot\overline{\_H},\cr
&\overline{\_B}=\overline{\overline{\zeta}}(\omega)\cdot\overline{\_E}+\overline{\overline{\mu}}(\omega)\cdot\overline{\_H},
\end{split}
\end{equation}
in which $\overline{\overline{\epsilon}}$, $\overline{\overline{\mu}}$, $\overline{\overline{\xi}}$, $\overline{\overline{\zeta}}$ are the frequency-domain material parameters. However, for a medium which is not stationary and its properties are time-variant, we need to express the constitutive relations which concurrently respect nonstationarity and memory. In the literature, assuming a time-varying dielectric isotropic medium ($\overline{\overline{\mu}}=\overline{\overline{\xi}}=\overline{\overline{\zeta}}=0$), that constitutive relation is often given by (e.g.,~Refs.~\cite{Halevieutv2009,Budko2009,Halevieutv2016})
\begin{equation}
\_D(t)=\epsilon(t)\_E(t).
\end{equation}
This model of a time-varying dielectric medium is based on a very dramatic approximation of instantaneous response of matter, which is not consistent with the temporal dispersion naturally present in materials. Therefore, more complete and rigorous definitions need to be introduced and applied. Indeed, for any point in space we should write (for an isotropic time-varying dielectric medium) \begin{equation}
\_D(t)=\epsilon_0\_E(t)+\int_0^{+\infty}\epsilon_0\chi(\gamma,t)\_E(t-\gamma)d\gamma,
\label{eq:mnopqadsus}
\end{equation}
in which $\chi(\gamma,t)$ is the electric susceptibility kernel (which may depend also on spatial coordinates).  In general,  time-varying fields can be expressed as an inverse Fourier transform
\begin{equation}
\_E(t)={1\over2\pi}\int_{-\infty}^{+\infty}\overline{\_E}(\omega)\exp(j\omega t)d\omega.    
\end{equation} 
Substituting the above expression in Eq.~\eqref{eq:mnopqadsus}, we deduce that
\begin{equation}
\_D(t)={\epsilon_0\over2\pi}\int_{-\infty}^{+\infty}\epsilon_{\rm{T}}(\omega,t)\overline{\_E}(\omega)\exp(j\omega t)d\omega,
\label{eq:DepsEomegat}
\end{equation}
where the temporal complex relative permittivity equals
\begin{equation}
\epsilon_{\rm{T}}(\omega,t)=1+\chi_{\rm{T}}(\omega,t)
\label{eq:repersusomegat}
\end{equation}
with
\begin{equation}
\chi_{\rm{T}}(\omega,t)=\int_0^{+\infty}\chi(\gamma,t)\exp(-j\omega\gamma)d\gamma.
\label{eq:susomegatF}
\end{equation}
This is similar to the definition used by N.~S.~Stepanov in Ref.~\cite{Stepanov1976} for describing macroscopic susceptibility of time-varying plasma. Here, the indices ``T'' are used to discern the temporal complex functions, $\epsilon_{\rm{T}}(\omega,t)$ and $\chi_{\rm{T}}(\omega,t)$, from the relative permittivity and susceptibility kernels, respectively (index ``T'' refers to ``temporal''). Based on the above derivations and explanations, what we unequivocally perceive is the fact that the temporal complex susceptibility and relative permittivity are general concepts which are  valid and useful even if the field is nonharmonic. Later, in the last part of Section~\ref{electron}, we employ these important equations and definitions, Eqs.~\eqref{eq:mnopqadsus}--\eqref{eq:susomegatF}, to calculate the effective macroscopic parameters of time-varying materials. 

In the end of this discussion, we should mention that in analogue to Eq.~\eqref{eq:2angfrerev}, we can take the Fourier transform of the electric flux density given by Eq.~\eqref{eq:DepsEomegat}. This operation significantly helps, allowing us to study and solve the Maxwell equations in the frequency domain ($\Omega$ domain). Keeping in mind that $\overline{\epsilon}_{\rm{T}}(\omega,\Omega)$ is the Fourier transform of $\epsilon_{\rm{T}}(\omega,t)$ with respect to the time variable $t$, we simply conclude that
\begin{equation}
\overline{\_D}(\Omega)={\epsilon_0\over2\pi}\int_{-\infty}^{+\infty}\overline{\epsilon}_{\rm{T}}(\omega,\Omega-\omega)\overline{\_E}(\omega)d\omega.
\label{eq:intintrigdeM}
\end{equation}
To appraise this relation, we carefully look at two particular instances. The first one is if the medium is immutable and stationary, but dispersive. Thus, the relative permittivity kernel depends on only one time variable and is given by $\epsilon(\gamma,t)=\epsilon(\gamma)$. Consequently, the temporal complex function and its Fourier transform are expressed as $\epsilon_{\rm{T}}(\omega,t)=\overline{\epsilon}(\omega)$ and $\overline{\epsilon}_{\rm{T}}(\omega,\Omega)=2\pi\overline{\epsilon}(\omega)\delta(\Omega)$, respectively, in which $\delta(\Omega)$ is the one-dimensional Dirac delta function. By plugging this result in Eq.~\eqref{eq:intintrigdeM} and using the property that $\int f(x)\delta(x-x_0)dx=f(x_0)$, we obtain $\overline{\_D}(\Omega)=\epsilon_0\overline{\epsilon}(\Omega)\overline{\_E}(\Omega)$ which is the conventional expression for modeling dispersive time-invariant media. For the second case, we assume that the medium possesses instantaneous response and varies in time. As we wrote earlier, this is what a multitude of research works have assumed in their studies of interactions of waves with time-varying media. Regarding this case, the relative permittivity kernel is $\epsilon(\gamma,t)=\delta(\gamma)\epsilon_{\rm{I}}(t)$ which results in $\epsilon_{\rm{T}}(\omega,t)=\epsilon_{\rm{I}}(t)$ and $\overline{\epsilon}_{\rm{T}}(\omega,\Omega)=\overline{\epsilon}_{\rm{I}}(\Omega)$. The initial observation, based on Eq.~\eqref{eq:DepsEomegat}, is that $\_D(t)=\epsilon_0\epsilon_{\rm{I}}(t)\_E(t)$, and the next observation, in accordance with Eq.~\eqref{eq:intintrigdeM}, explains the fact that $\overline{\_D}(\Omega)$ is the convolution of $\overline{\epsilon}_{\rm{I}}(\Omega)$ and the field $\overline{\_E}(\Omega)$. Both these two observations are quite expected for the nondispersive model of time-varying media.


\section{Individual time-varying particle in free space}
\label{Appendix:nsdpktcp}
Based on the fundamental notions introduced and discussed above, we can address an important question on  how to determine the polarizability kernel $\alpha(\gamma,t)$ and the temporal complex polarizability $\alpha_{\rm{T}}(\omega,t)$ for a single time-varying particle located in free space. To answer this central question, first, we need to write a linear differential equation which describes the polarization dynamics of time-varying particles. Studying this equation and using the fundamentals explained before, we will develop a systematic method to find the particle  polarizability.

Here, let us focus on the canonical example of a nonstationary Hertzian dipole. The understanding of this basic scatterer can be extended to small time-varying inclusions which have electric response. We model a Hertzian dipole as a short wire antenna of length $l$, assuming that the current is uniform along the wire. The antenna parameters are its effective inductance $L$, capacitance $C$, and a  resistive load $R$ (that accounts for dissipation losses). Parameters $L$ and $C$  measure the magnetic and electric energies stored in the reactive fields around the dipole. Accordingly, for such a Hertzian dipole, a third-order differential equation for the oscillating charge $Q$ is expressed as
\begin{equation}
-Z{d^3Q\over dt^3}+L{d^2Q\over dt^2}+R{dQ\over dt}+{1\over C}Q=lE.
\label{EQ:RUD1907}
\end{equation}
This is the ``R{\"u}denberg equation" which was initially written in 1907 for elucidating the Hertzian dipole antenna in the receiving regime~\cite{rudenberg1907}. On the right side, $E$ denotes the incident electric field at the dipole location, parallel to the dipole. Notice that the first term in the above equation on the left side, which is proportional to the third time derivative of the electric charge, is associated with the radiation of the dipole and is linked with the radiation reaction (see, e.g., Ref.~\cite{TM-Radiation2}). The parameter $Z=l^2\mu_0/(6\pi c)$ determines what R{\"u}denberg calls ``Strahlungswiderstand" or radiation resistance (here, $\mu_0$ is the free-space permeability, and $c$ denotes the speed of light). Including this first term in Eq.~\eqref{EQ:RUD1907}, we see that the R{\"u}denberg equation is in fact analogous to the Abraham-Lorentz equation~\cite{NH2012} that also includes the radiation reaction term. Since the electric dipole moment is the multiplication of the electric charge and the dipole length, therefore, we readily modify the authentic version of R{\"u}denberg equation and write
\begin{equation}
-l{\mu_0\over6\pi c}{d^3\_p\over dt^3}+{L\over l}{d^2\_p\over dt^2}+{R\over l}{d\_p\over dt}+{1\over lC}\_p=l\_E.
\label{appfirwlotvns}
\end{equation}
Till this point, we have assumed that the Hertzian dipole is stationary and the resistance, inductance, and the capacitance are immutable and constant in time. To realize a  nonstationary Hertzian dipole, it is sufficient to make those parameters time-dependent in the differential equation. For instance, suppose that we add a time-varying capacitance as an extra load which is connected in series with a dynamic load resistance $R(t)$. For such time-varying dipoles, the differential equation, Eq.~\eqref{appfirwlotvns}, is rewritten as
\begin{equation}
-{l^2\over L}{\mu_0\over6\pi c}{d^3\_p\over dt^3}+{d^2\_p\over dt^2}+{R(t)\over L}{d\_p\over dt}+{1\over LC(t)}\_p={l^2\over L}\_E.
\label{appnxsubBtvhd}
\end{equation}
Note that $C(t)$ is the total capacitance which contains both the effective capacitance due to the stored electric energy around the dipole and the time-modulated load capacitance. Also, note that in Eq.~\eqref{appnxsubBtvhd}, if the first term (the radiation term) on the left side is neglected (by assuming that it is much smaller than the last term (the restoring-force term)), we achieve the differential equation that is utterly similar to the equation of motion for bound electrons in matter, discussed in the next section (see Eq.~\eqref{eq:drloreq}). Here, however, we study a dipole in free space and keep the radiation term. 

Equation~\eqref{appnxsubBtvhd} relates the instantaneous dipole moment and the excitation field. On the other hand, in Section~\ref{sec:basicconcepts}, we related the instantaneous dipole moment and the field in terms of the polarizability kernel. Therefore, by bringing these two models together, we can find the polarizability kernel (and the temporal complex polarizability) in terms of the time-varying resistance and capacitance. We start from Eq.~\eqref{causaleq:dp}, which is an alternative integral form to describe the dipole moment $\_p$ of a time-varying particle excited by an external electric field $\_E$: $\_p(t)=\int_{-\infty}^t h(t,\tau)\_E(\tau)d\tau$. Next, we need to replace this description in Eq.~\eqref{appnxsubBtvhd}. For replacing, however, we have to calculate the first, second, and third time derivatives of the dipole moment. To find these derivatives, we use the chain rule and the Leibniz integral rule which say that for any integrable function $f(x,y)$, we can write 
\begin{equation}
\begin{split}
&{d\over dx}\int_{a(x)}^{b(x)}f(x,y)dy=\cr
&f(x,b(x)){db(x)\over dx}-f(x,a(x)){da(x)\over dx}+\int_{a(x)}^{b(x)}{\partial\over\partial x}f(x,y)dy,
\end{split}
\label{eqLinruLinabx}
\end{equation}
where $a(x)$ and $b(x)$ denote the lower and upper limits, respectively. By employing Eq.~\eqref{eqLinruLinabx} and doing careful algebraic manipulations, the first, second, and third derivatives of the dipole moment are written in terms of $h(t,\tau)$ and the corresponding partial derivatives of $h(t,\tau)$ as
\begin{equation}
\begin{split}
&{d\_p\over dt}=h(t,\tau)\vert_{\tau=t}\_E+\int_{-\infty}^t{\partial h(t,\tau)\over\partial t}\_E(\tau)d\tau,\cr
&{d^2\_p\over dt^2}=\Bigg[2{\partial h(t,\tau)\over\partial t}\vert_{\tau=t}+{\partial h(t,\tau)\over\partial\tau}\vert_{\tau=t}\Bigg]\_E\cr
&+h(t,\tau)\vert_{\tau=t}{d\_E\over dt}+\int_{-\infty}^t{\partial^2h(t,\tau)\over\partial t^2}\_E(\tau)d\tau,
\end{split}
\label{eq:fdpsdeinpol}
\end{equation} 
and
\begin{equation}
\begin{split}
&{d^3\_p\over dt^3}=\Bigg[3{\partial^2h(t,\tau)\over\partial t^2}+3{\partial^2h(t,\tau)\over\partial t\partial\tau}+{\partial^2h(t,\tau)\over\partial\tau^2}\Bigg]_{\tau=t}\_E+\cr
&\Bigg[3{\partial h(t,\tau)\over\partial t}+2{\partial h(t,\tau)\over\partial\tau}\Bigg]_{\tau=t}{d\_E\over dt}+h(t,\tau)\vert_{\tau=t}{d^2\_E\over dt^2}+\cr
&\int_{-\infty}^t {\partial^3h(t,\tau)\over\partial t^3}\_E(\tau)d\tau.
\end{split}
\label{eq:appendixdiptvC}
\end{equation}
Ultimately, by substituting the results of Eqs.~\eqref{eq:fdpsdeinpol} and \eqref{eq:appendixdiptvC} into Eq.~\eqref{appnxsubBtvhd}, we derive four expressions (one characteristic equation and three initial conditions) which define the function $h(t,\tau)$. These four expressions read 
\begin{equation}
\begin{split}
&1)\,\,\,\mathbb{R}{\partial^3h(t,\tau)\over\partial t^3}+{\partial^2h(t,\tau)\over\partial t^2}+{R\over L}{\partial h(t,\tau)\over\partial t}+\cr 
&\,\,\,\,\,\,\,\,{1\over LC(t)}h(t,\tau)=0,\cr
&2)\,\,\,3\mathbb{R}{\partial^2h(t,\tau)\over\partial t^2}\vert_{\tau=t}+3\mathbb{R}{\partial^2h(t,\tau)\over\partial t\partial\tau}\vert_{\tau=t}+\cr
&\,\,\,\,\,\,\,\,\mathbb{R}{\partial^2h(t,\tau)\over\partial\tau^2}\vert_{\tau=t}+2{\partial h(t,\tau)\over\partial t}\vert_{\tau=t}+{\partial h(t,\tau)\over\partial\tau}\vert_{\tau=t}={l^2\over L},\cr
&3)\,\,\,3{\partial h(t,\tau)\over\partial t}\vert_{\tau=t}+2{\partial h(t,\tau)\over\partial\tau}\vert_{\tau=t}=0,\cr
&4)\,\,\,h(t,\tau)\vert_{\tau=t}=0,
\end{split}
\label{eq:APP3DerCAP}
\end{equation}
in which $\mathbb{R}=-(l^2/L)(\mu_0/6\pi c)$.

At this point, one may ask what will happen if the dipole is loaded with a time-varying inductance rather than a time-varying capacitance. How do the above expressions change? This is a valid and intriguing question. In fact, we should first revise the third-order differential equation which  describes the nonstationary dipole connected to  time-varying inductance. For that, as one may expect, we start from the primary version of the R{\"u}denberg equation, Eq.~\eqref{EQ:RUD1907},  which is based on the electromotive force, and we rewrite the voltage over the time-varying inductance in this equation. Subsequently, after a simple modification, we express the desired differential equation that is similar to Eq.~\eqref{appnxsubBtvhd}. Therefore, we have 
\begin{equation}
-{l^2\over L(t)}{\mu_0\over6\pi c}{d^3\_p\over dt^3}+{d^2\_p\over dt^2}+\Gamma(t){d\_p\over dt}+{1\over L(t)C}\_p={l^2\over L(t)}\_E,
\label{eq:TVINDDIPSec3}
\end{equation} 
where $\Gamma(t)=[1/L(t)][R+dL(t)/dt]$ is a function that is related to the time derivative of the inductance. Consequently, this function becomes a constant if the derivative of the inductance vanishes. Here, we do not continue with the derivations of the polarizability kernel because the method for deriving them is quite clear for the readers (Eqs.~\eqref{eq:fdpsdeinpol} and \eqref{eq:appendixdiptvC} should be substituted into Eq.~\eqref{eq:TVINDDIPSec3}). Therefore, we pass those derivations and calculations on to the interested reader. 

The above theory, which we developed for determining the polarizability kernel, will be complemented by deriving another partial differential equation that ``directly" describes the temporal complex polarizability when the incident electric field is time-harmonic. Similar to what we did for the polarizability kernel  in Eq.~\eqref{appnxsubBtvhd}, it is sufficient to supersede the dipole moment by its definition in terms of $\alpha_{\rm{T}}(\omega,t)$ (i.e.,~$\_p(t)={\rm{Re}}\big[\alpha_{\rm{T}}(\omega,t)\cdot\_E_0\exp(j\omega t)\big]$) in order to derive such a partial differential equation. After doing simplifications, we write that
\begin{equation}
\begin{split}
&\mathbb{R}{\partial^3\alpha_{\rm{T}}(\omega,t)\over\partial^3t}+\bigg(1+j3\omega\mathbb{R}\bigg){\partial^2\alpha_{\rm{T}}(\omega,t)\over\partial^2t}+\cr
&\bigg({R\over L}+j2\omega-3\omega^2\mathbb{R}\bigg){\partial\alpha_{\rm{T}}(\omega,t)\over\partial t}+\cr
&\bigg({1\over LC(t)}-\omega^2+j\omega{R\over L}-j\omega^3\mathbb{R}\bigg)\alpha_{\rm{T}}(\omega,t)={l^2\over L}.
\end{split}
\end{equation}
Notice that here we assumed that the nonstationary dipole is loaded by a time-varying capacitance. However, one can repeat the same procedure when the dipole load is a time-varying inductance (see  Eq.~\eqref{eq:TVINDDIPSec3}).


\subsection{Dipole interaction with incident waves}
\label{secHertziandipole}

The analytical approach for determining the polarizability kernel $\alpha(\gamma,t)$ and the temporal complex polarizability $\alpha_{\rm{T}}(\omega,t)$ of  nonstationary dipoles was explained in the previous part of this section. At the following step, we scrutinize classical interactions of nonstationary dipoles with external fields based on the notion of polarizability, and study instantaneous powers exerted on and radiated by the dipole under illumination. 

\subsubsection{Instantaneous power exerted on dipole} 
Let us assume that a Hertzian dipole is located at the origin of the Cartesian coordinate system, directed along the $z$ axis, and illuminated by a time-harmonic electric field. Also, let us assume that the nonstationary characteristic of the dipole can be realized, for example, by loading the dipole with a time-varying lumped element ~\cite{TM-Radiation1} or varying the dipole length in time. For the first case, in which the effective length of the Hertzian dipole is fixed and the dipole is  loaded by a time-varying lumped element, we can simply employ the concept of induced electromotive force and calculate the total instantaneous power exerted on the dipole. From the basics, we know that the electric dipole moment is  the multiplication of the electric charge $Q(t)$ and the dipole length $l$: $p(t)=lQ(t)$. Since the time derivative of the electric charge is identical with the electric current, as a result, the time derivative of the dipole moment becomes equal to the length of the dipole $l$ multiplied by the electric current $i(t)$ carried by the dipole: $dp(t)/dt=l\cdot i(t)$. We stress that in this simple calculation, the length of the dipole is not changing in time, and the temporal variation of the dipole moment is only due to the temporal variation in the electric charge. On the other hand, the induced electromotive force corresponding to effective length $l$ is expressed as $v(t)=l\cdot E(t)$, in which $E(t)$ is the component of the excitation field parallel to the Hertzian dipole. Using the above two equations, representing the induced electromotive force and the time derivative of the dipole moment, the instantaneous power is obtained from $S(t)=v(t)\cdot i(t)$, which finally gives
\begin{equation}
S(t)=E(t)\cdot{dp(t)\over dt}. 
\label{eq:elecdipolemulti}
\end{equation}
Regarding the second case, where the length of the dipole is also changing in time, and, therefore, the temporal variation of the dipole moment is not only due to the electric charge $Q(t)$, but it is also due to the length variations $l(t)$, it may initially seem to be complicated to find the instantaneous power. However, in general, the electric current density corresponding to the Hertzian dipole is written as $\_J(\_r,t)={d\_p(t)\over dt}\delta^3(\_r)$, and the instantaneous power $S(t)$ is expressed as 
\begin{equation}
S(t)=\int_V\_J\cdot\_E\;d^3\_r,
\label{eq:genecdtsn}
\end{equation}where $V$ denotes the volume occupied by the dipole. Based on Eq.~\eqref{eq:genecdtsn} and by substituting the electric current density, we explicitly achieve the same expression as given by Eq.~\eqref{eq:elecdipolemulti}. Therefore, we conclude that Eq.~\eqref{eq:elecdipolemulti} is valid even for the case when the dipole length is changing in time. Notice that Eq.~\eqref{eq:elecdipolemulti} is also true for any time variation of the dipole moment, including the stationary scenario.

According to Eq.~\eqref{eq:dmpolth}, the dipole moment $p(t)$ is described in terms of the temporal complex polarizability $\alpha_{\rm{T}}(\omega,t)$. Therefore, this equation can be substituted into Eq.~\eqref{eq:elecdipolemulti} in order to  present the power $S(t)$ in terms on the polarizability $\alpha_{\rm{T}}(\omega,t)$. Before we proceed and do the substitution, let us note that due to the single-frequency excitation, for brevity, we can drop the first argument of $\alpha_{\rm{T}}(\omega,t)$ (i.e.~the angular frequency of the incident field). In addition, for simplicity, we also define the following complex-valued function:
\begin{equation}
\zeta(t)=\alpha_{\rm{T}}(t)+{1\over j\omega}{d\alpha_{\rm{T}}(t)\over dt},
\label{eq:chimodp}
\end{equation}
which is associated with the temporal complex polarizability and its time derivative. Now, by substituting the temporal complex polarizability and using this auxiliary function definition, Eq.~\eqref{eq:chimodp}, the extracted power is simplified to 
\begin{equation}
S(t)=E(t)\cdot\Re\Big[j\omega\zeta(t)\cdot E_0\exp(j\omega t)\Big].
\end{equation}
Writing the real part as $\Re[x]=(1/2)(x+x^\ast)$, finally, the extracted power reduces to 
\begin{equation}
S(t)=-{\omega\over2}\Im\big[\zeta(t)\big]\vert E_0\vert^2-{\omega\over2}\Im\bigg[\zeta(t)E^2_0\exp(j2\omega t)\bigg],
\label{eq:mainext}
\end{equation}
in which ${\rm{Im}}[\,\,]$ denotes the imaginary part.

Let us check this equation for the special case of a stationary dipole. In this case the time derivative of $\alpha_{\rm{T}}(t)$ vanishes, and $\zeta$ becomes a complex constant which is equal to $\alpha_{\rm{T}}$. As a consequence, the time-averaged value of the second term in the above equation becomes zero and the time-averaged power extracted by the dipole from the incident field is simply  $\overline{S^{\rm{stationary}}}=-{\omega\over2}\Im[\zeta]\vert E_0\vert^2=-{\omega\over2}\Im[\alpha_{\rm{T}}]\vert E_0\vert^2$, which is the same relation as we know from the literature (see e.g.~Ref.~\cite{Tretyakovplasmonics}). Here, we stress that for stationary dipoles the expression in Eq.~\eqref{eq:mainext} is also valid in the time domain. 

Another special case is the case when $\zeta(t)=0$. From Eq.~\eqref{eq:mainext} we see that if $\zeta(t)=0$, the extracted power $S(t)$ is zero meaning that the dipole does not interact with the incident field. According to Eq.~\eqref{eq:chimodp}, the condition $\zeta(t)=0$ corresponds to $\alpha_{\rm{T}}(t)=Ae^{-j\omega t}$ where $A$ is a constant coefficient. Substituting this result into Eq.~\eqref{eq:dmpolth}, we see that the dipole moment is constant over time. In other words, we have a static dipole moment whose time derivative is zero. Consequently, there should not be any interaction with the incident field. 

Considering Eq.~\eqref{eq:mainext}, it is intriguing to assume a periodic function $\zeta(t)$. This is because periodicity allows us to employ simple time averaging. Based on the Fourier series written for a periodic function, depending on the period and the complex Fourier coefficients, the time-averaged value  associated with the second term in Eq.~\eqref{eq:mainext} is not zero,  and it can significantly contribute to the time-averaged total power. For example, it is clearly seen that if the period is equal to the excitation period $T=2\pi/\omega$, the second-order term in the Fourier series $n=2$ can produce a nonzero averaged value (in contrast with the stationary case, in which the average is zero). 

\subsubsection{Instantaneous power radiated by dipole}
Some part of the extracted power is  re-radiated to the background medium (here, free space). The instantaneous power which is re-radiated by the dipole is proportional to the first and the third time derivatives of the dipole moment~\cite{TM-Radiation1,TM-Radiation2}: 
\begin{equation}
S_{\rm{rad}}(t)=-{\mu_0\over6\pi c}{dp(t)\over dt}\cdot{d^3p(t)\over dt^3}.
\label{eq:radpowscat}
\end{equation}
Similarly to what we did for $S(t)$, we substitute the temporal complex polarizability in the above equation for the re-radiated power. In accordance with Eqs.~\eqref{eq:dmpolth} and \eqref{eq:chimodp}, the first and the third time derivatives of the dipole moment are given by
\begin{equation}
\begin{split}
&{dp(t)\over dt}=\Re\Bigg[j\omega\zeta(t)\cdot E_0\exp(j\omega t)\Bigg],\cr
&{d^3p(t)\over dt^3}=\Re\Bigg[j\omega\Big({d^2\zeta\over dt^2}+j2\omega{d\zeta\over dt}-\omega^2\zeta\Big)\cdot E_0\exp(j\omega t)\Bigg].
\end{split}
\end{equation}
Subsequently, by using these equations and after doing some algebraic manipulations, we find the re-radiated power (in Eq.~\eqref{eq:radpowscat}) as
\begin{equation}
\begin{split}
&S_{\rm{rad}}(t)={\mu_0\omega^4\over12\pi c}\vert\zeta\vert^2\vert E_0\vert^2+\cr
&{\mu_0\omega^4\over12\pi c}\vert E_0\vert^2\Re\Bigg[\zeta\bigg({1\over\omega^2}{d^2\zeta\over dt^2}+j{2\over\omega}{d\zeta\over dt}-\zeta\Big){E_0^2\over\vert E_0\vert^2}\exp(j2\omega t)\cr
&-\zeta\bigg({1\over\omega^2}{d^2\zeta\over dt^2}+j{2\over\omega}{d\zeta\over dt}\bigg)^\ast\Bigg].
\end{split}
\end{equation}
If the dipole is stationary, all the time derivatives in the above equation become zero, and the time-averaged scattered power is simplified to
$\overline{S^{\rm{stationary}}_{\rm{rad}}}={\mu_0\omega^4\over12\pi c}\vert\zeta\vert^2\vert E_0\vert^2={\mu_0\omega^4\over12\pi c}\vert\alpha_{\rm{T}}\vert^2\vert E_0\vert^2$,
which can be conveniently found in the literature (e.g.~Ref.~\cite{Tretyakovplasmonics}). 

To summarize, for a nonstationary electric dipole under illumination, we first derived the corresponding differential equations (with the corresponding initial conditions) for obtaining the polarizability. Afterwards, we utilized the introduced notion of the temporal complex polarizability to find the total instantaneous extracted power and the instantaneous scattered power, which are not necessarily equal to each other (see Appendix~\ref{appeninvpolinsrpow} for more information).


\section{Nonstationary particle as a constituent of a time-varying material} 
\label{electron}

In contrast to the previous section, where the time-varying particle is located in free space, in this section, we assume that the particle is a constituent of a time-varying material, and, accordingly, we investigate the polarizability of such a particle. This investigation is important for understanding the effective macroscopic parameters such as susceptibility and permittivity of  dynamic materials. To determine the polarizability kernel and the temporal complex polarizability, we need to study the corresponding differential equation. Since the particle is immersed in a time-varying material composed of many identical particles, the radiated power is compensated by the power received from other particles. This cancellation ensures that in the absence of dissipation in the particles and power exchange with the external devices that modulate the particles, the effective medium parameters correspond to a lossless material. In this case, the order of the differential equation is two because the radiation reaction term that was proportional to the third derivative of the dipole moment vanishes.

\subsection{Differential equations for the dipole moment and the polarizability kernel}
Let us concentrate on the classical model of a bound electron as our particle under study. In this case, the  differential equation needed for  description of the electric dipole moment is given by the classical equation of motion. Based upon this equation, we model external time modulations of the system by assuming that the damping coefficient $\Gamma_{\rm{D}}$ and the natural frequency $\omega_{\rm{n}}$ are varying in time. Therefore, the second-order differential equation describing the electron motion is expressed as ${d^2x(t)\over dt^2}+\Gamma_{\rm{D}}(t){dx(t)\over dt}+\omega_{\rm{n}}^2(t)x(t)={e\over m}E(t)$, in which $m$ denotes the electron mass, $e$ represents the electron charge, and $x(t)$ is the displacement. Since the dipole moment is the multiplication of the electron charge and the displacement, we can consequently write  
\begin{equation}
{d^2\_p\over dt^2}+\Gamma_{\rm{D}}(t){d\_p\over dt}+\omega_{\rm{n}}^2(t)\_p={e^2\over m}\_E.
\label{eq:drloreq}
\end{equation}
Here, it is worth noting that one can entitle Eq.~\eqref{eq:drloreq} as the Lorentz equation which results in the Lorentz model for the effective macroscopic parameters of dielectric materials. This model reduces to the Drude model in the limit of vanishing natural frequency ($\omega_{\rm{n}}=0)$, which means that there is no restoring  force and the electron is not bound. Therefore, in general, we use the term ``Drude-Lorentz'' to entitle Eq.~\eqref{eq:drloreq}.
By employing Eqs.~\eqref{causaleq:dp} and \eqref{eqLinruLinabx}, we already derived the first and second time derivatives of the electric dipole moment and demonstrated them based on the function $h(t,\tau)$ and its partial derivatives. Thus, by inserting those derivations (written in Eq.~\eqref{eq:fdpsdeinpol}) into Eq.~\eqref{eq:drloreq} (the Drude-Lorentz equation), we arrive to three crucial expressions which determine the polarizability:
\begin{equation}
\begin{split}
&1)\,\,\,\,\,\,\,\,{\partial^2h(t,\tau)\over\partial t^2}+\Gamma_{\rm{D}}(t){\partial h(t,\tau)\over\partial t}+\omega_{\rm{n}}^2(t)h(t,\tau)=0,\cr
&2)\,\,\,\,\,\,\,\,2{\partial h(t,\tau)\over\partial t}\vert_{\tau=t}+{\partial h(t,\tau)\over\partial\tau}\vert_{\tau=t}={e^2\over m},\cr
&3)\,\,\,\,\,\,\,\,h(t,\tau)\vert_{\tau=t}=0.
\end{split}
\label{eq:mtmio}
\end{equation} 
We stress that the function $h(t,\tau)$ is not the polarizability kernel. Indeed, the polarizability kernel that we introduced above is $\alpha(\gamma,t)=h(t,\tau)$ when $\tau=t-\gamma$. As a consequence, $\tau=t$ in the second and third expressions refers to $\gamma=0$, and Eq.~\eqref{eq:mtmio} defines two initial conditions at  $\gamma=0$ for $\alpha(\gamma,t)$. Depending on the temporal functions of the damping coefficient and the natural frequency, these three expressions in Eq.~\eqref{eq:mtmio} give a specific function for the polarizability. 

As a check, let us examine the results by considering first the stationary scenario, assuming that the damping coefficient and the natural frequency are not varying in time. By remembering that $h(t,\tau)=h(t-\tau)$ in this scenario and solving Eq.~\eqref{eq:mtmio}, the polarizability kernel is derived as
\begin{equation}
\begin{split}
&\alpha(\gamma,t)=h(t,t-\gamma)=\cr
&{e^2\over m\sqrt{\omega_{\rm{n}}^2-{\Gamma_{\rm{D}}^2\over4}}}\exp\big(-{\Gamma_{\rm{D}}\over2}\gamma\big)\sin\big(\sqrt{\omega_{\rm{n}}^2-{\Gamma_{\rm{D}}^2\over4}}\gamma\big).
\end{split}
\label{appendixfullderi}
\end{equation}
Notice that the full derivation is explained in the Appendix of the paper. In the above equation, as it is explicitly seen, the polarizability kernel depends on only one time parameter, $\gamma$. Therefore, its Fourier transform is only a function of the frequency and gives the known Drude-Lorentz dispersion in the frequency domain~(e.g., \cite{jackson}). 

Regarding the nonstationary scenario, in which the damping coefficient or the natural frequency depends on time, we will give a complete example in the last part of this section,  comprehensively calculating the polarizability kernel (see Eqs.~\eqref{eq:httfullexamplernsm}--\eqref{eq:polartemptVdamping} and the corresponding explanations). Also, subsequently, we will compare the obtained results with the ones known for the conventional stationary scenario. In this example, we will assume that the damping coefficient is temporally modulated as $\Gamma_{\rm{D}}(t)=2\Gamma_0/(1+\Gamma_0t)$ (where $\Gamma_0$ corresponds to the damping coefficient at $t=0$) and the natural frequency is zero.

\subsubsection{Noncausal interpretation}
Prior to studying the temporal complex polarizability, here, we would like to have a brief discussion around causality of  time-modulated particles. If we do not respect causality, the dipole moment can also depend on the electric field in the future, and, therefore, $\_p(t)=\int_{-\infty}^{+\infty}h(t,\tau)\_E(\tau)d\tau$. Such interpretation affects strikingly the results of Leibniz integral expressions. In fact, by applying Eq.~\eqref{eqLinruLinabx}, the first and second derivatives of the dipole moment are modified as 
\begin{equation}
\begin{split}
&{d\_p\over dt}=\int_{-\infty}^{\infty}{\partial h(t,\tau)\over\partial t}\_E(\tau)d\tau,\cr
&{d^2\_p\over dt^2}=\int_{-\infty}^{\infty}{\partial^2h(t,\tau)\over\partial t^2}\_E(\tau)d\tau.
\end{split}
\label{EQNcNcNcNcINT}
\end{equation}
We can compare these equations with the expressions for the causal interpretation (Eq.~\eqref{eq:fdpsdeinpol}), in order to understand the fundamental difference between them. By using Eq.~\eqref{EQNcNcNcNcINT} and considering Eq.~\eqref{eq:drloreq}, which is the Drude-Lorentz equation, we write 
\begin{equation}
\begin{split}
&\int_{-\infty}^{\infty}\Bigg[{\partial^2h(t,\tau)\over\partial t^2}+\Gamma_{\rm{D}}(t){\partial h(t,\tau)\over\partial t}+\omega_{\rm{n}}^2(t)h(t,\tau)\Bigg]\_E(\tau)d\tau\cr
&={e^2\over m}\_E(t).
\end{split}
\end{equation}
As we explicitly see from the above, since the electric field is simultaneously attending both sides of the equality and it is inside the integral on the left side, we can conclude that the whole expression within the square brackets should be equal to the Dirac delta function. In other words,
\begin{equation}
{\partial^2h(t,\tau)\over\partial t^2}+\Gamma_{\rm{D}}(t){\partial h(t,\tau)\over\partial t}+\omega_{\rm{n}}^2(t)h(t,\tau)={e^2\over m}\delta(\tau-t).  \label{eqmohkhncs}
\end{equation}
Equation~\eqref{eqmohkhncs} is the key equation for calculating the polarizability kernel describing a noncausal response. Solving this equation and finding a solution for the function $h(t,\tau)$ may not be straightforward (and even possible) due to the presence of the Dirac delta function on the right side. However, if there is a nonzero solution, notice that it must be a real solution in time because the function $h(t,\tau)$ is indeed a real-valued  function. Here, we also point out an intriguing issue which influences Eq.~\eqref{eqmohkhncs}. In the  expression $\_p(t)=\int_{-\infty}^{+\infty}h(t,\tau)\_E(\tau)d\tau$ written initially, although we chose the upper limit of the integral as infinity, we can definitely assume a finite upper limit and still describe a noncausal response. The only condition is that the finite upper limit should be larger than the observation time $t$, i.e.~${\rm{upper\,limit}}=t+\beta$ where $\beta$ is real and $\beta>0$. In accordance with this assumption, $\_p(t)=\int_{-\infty}^{t+\beta}h(t,\tau)\_E(\tau)d\tau$, subsequently, we can simply rewrite Eq.~\eqref{eqmohkhncs} by using the Leibniz integral rule and Drude-Lorentz equation. In this paper, since our focus is only on the causal response, we leave such derivations for the interested reader.


\subsection{Differential equation for finding the temporal complex polarizability} 

We have hitherto discussed how to derive the polarizability kernel of the electron based on the classical model. Here,  our aim is to introduce a linear differential equation whose solution gives the temporal complex polarizability. We already know that, for time-harmonic excitation, the induced electric dipole moment is expressed in terms of $\alpha_{\rm{T}}(\omega,t)$ (see Eq.~\eqref{eq:dmpolth}). Ergo, by substituting that expression into the Drude-Lorentz equation (see Eq.~\eqref{eq:drloreq}), we come to the desired differential equation for the temporal complex polarizability, which is written as
%
%
\begin{equation}
\begin{split}
{\partial^2\alpha_{\rm{T}}(\omega,t)\over\partial t^2}+&\Big(\Gamma_{\rm{D}}(t)+j2\omega\Big)
{\partial\alpha_{\rm{T}}(\omega,t)\over\partial t}+\cr
&\Big(\omega^2_{\rm{n}}(t)-\omega^2+j\omega \Gamma_{\rm{D}}(t)\Big)\alpha_{\rm{T}}(\omega,t)={e^2\over m}.
\end{split}
\label{EQEQ:IMPORTANTTCPOL}
\end{equation}
Equation~\eqref{EQEQ:IMPORTANTTCPOL} is a second-order differential equation with time-dependent coefficients which allows us to find $\alpha_{\rm{T}}(\omega,t)$ for arbitrary time variations of the particle parameters. Certainly, this equation should be complemented by initial conditions in order to obtain a specific solution for $\alpha_{\rm{T}}(\omega,t)$. These initial conditions are given by Eq.~\eqref{eq:mtmio} in which $h(t,\tau)$ and the polarizability kernel are present. Therefore, after solving Eq.~\eqref{EQEQ:IMPORTANTTCPOL} and finding $\alpha_{\rm{T}}(\omega,t)$, one needs to firstly make the inverse Fourier transform to calculate the polarizability kernel $\alpha(\gamma,t)$ and subsequently $h(t,\tau)$. As a result, by having the function $h(t,\tau)$, the initial conditions expressed in Eq.~\eqref{eq:mtmio} (the second and third expressions) can be readily checked to achieve the specific solution.

In order to check Eq.~\eqref{EQEQ:IMPORTANTTCPOL}, we first make $\Gamma_{\rm{D}}(t)$ and $\omega_{\rm{n}}(t)$ time-invariant. Since $\alpha_{\rm{T}}(\omega,t)$ does not depend on time in this case, the corresponding time derivatives in Eq.~\eqref{EQEQ:IMPORTANTTCPOL} become zero, and we instantly see that the solution is the usual Lorentz dispersion rule: 
\begin{equation}
\alpha_{\rm{T}}(\omega,t)={{e^2\over m}\over\omega^2_{\rm{n}}-\omega^2+j\omega\Gamma_{\rm{D}}}.
\end{equation}
As a second check, we consider a time-variant $\Gamma_{\rm{D}}(t)$ or $\omega_{\rm{n}}(t)$. For this, let us take the same example as in the next subsection of the paper, where we will assume that the damping coefficient is $\Gamma_{\rm{D}}(t)=2\Gamma_0/(1+\Gamma_0t)$ and the natural frequency is zero. For this example, we apply Eq.~\eqref{eq:mtmio} and derive  the polarizability kernel which is given by Eq.~\eqref{eq:polartemptVdamping} in the next subsection. Therefore, since we know $\alpha(\gamma,t)$, we can obtain the temporal complex polarizability $\alpha_{\rm{T}}(\omega,t)$ by simply taking the Fourier transform with respect to $\gamma$, see  Eq.~\eqref{mostimeq}. After some algebraic manipulations, $\alpha_{\rm{T}}(\omega,t)$ is expressed as
\begin{equation}
\alpha_{\rm{T}}(\omega,t)=-{e^2\over m}\Big({1\over\omega^2}+j{2\over\omega^3}{\Gamma_0\over1+\Gamma_0t}\Big). 
\label{eq:tcppphopage}
\end{equation}
Now, as one can expect, this expression in Eq.~\eqref{eq:tcppphopage} should definitely satisfy the second-order differential equation  Eq.~\eqref{EQEQ:IMPORTANTTCPOL}. If we substitute $\alpha_{\rm{T}}(\omega,t)$ (written above) into Eq.~\eqref{EQEQ:IMPORTANTTCPOL}, we  observe that Eq.~\eqref{EQEQ:IMPORTANTTCPOL} indeed  holds. Notice that Eq.~\eqref{eq:tcppphopage} is not calculated based on Eq.~\eqref{EQEQ:IMPORTANTTCPOL}, but it is obtained by employing the comprehensive Eq.~\eqref{eq:mtmio}. Therefore, $\alpha_{\rm{T}}(\omega,t)$ in Eq.~\eqref{eq:tcppphopage} represents the general solution of the second-order differential equation, which consists of both the complementary function (the solution of the homogeneous equation) and the particular integral solution (the solution of the inhomogeneous equation). This general solution also respects the initial conditions expressed in Eq.~\eqref{eq:mtmio}. From this point of view, such general solution can be finally considered as the unique solution to Eq.~\eqref{EQEQ:IMPORTANTTCPOL} with $\omega_{\rm{n}}(t)=0$ and $\Gamma_{\rm{D}}(t)=2\Gamma_0/(1+\Gamma_0t)$.

\subsection{On the Drude-Lorentz model of time-varying dielectrics and plasma}

Next, we use the above theoretical results to analyse approximate models of effective parameters of Lorentzian dielectrics and electron plasma. The dipole moment of each electron is governed by Eq.~\eqref{eq:drloreq}, where the parameters may depend on time due to changing environment where the charges are located. However, apart from Eq.~\eqref{eq:drloreq}, the electron density (i.e.~the number of electrons per unit volume $N(t)$) can also depend on time. In consequence, in the following, we will consider two different cases: A particular case in which only the electron density is time variant and the parameters in the equation of motion (Eq.~\eqref{eq:drloreq}) do not change in time, and a more general case in which those parameters are also time dependent in addition to the electron density. Both aforementioned cases definitely result in nonstationary models. The reason for having a discussion on the former case is that while the polarizability kernel of the single electron should be the same as the one for the stationary scenario, because the damping coefficient and the natural frequency are assumed to be time invariant, and, therefore, the polarizability kernel is only a function of $\gamma$ (i.e.~$\alpha(\gamma,t)=\alpha(\gamma)$), we will soon show that the corresponding effective permittivity kernel becomes a function of both $\gamma$ and $t$ (i.e.~$\epsilon(\gamma,t)$).

Accordingly, let us start from the first case which is possibly the simplest case. Since $\Gamma_{\rm{D}}$ and $\omega_{\rm{n}}$ are constant in time and only the density $N(t)$ varies, we can see this as a low-density approximation where we assume that the electrons interact very weakly and, as a result, the characteristics of movement of a single electron do not depend on the electron density. Under these assumptions, the volume density of electric dipole moment or polarization density is written as \begin{equation}
\begin{split}
\mathbf{P}(t)=N(t)\mathbf{p}(t)=&\int_{0}^{+\infty}N(t)\alpha(\gamma,t)\_E(t-\gamma)d\gamma=\cr
&\int_{0}^{+\infty}\varepsilon_0\chi(\gamma,t)\_E(t-\gamma)d\gamma,
\end{split}
\label{eq:pdntptsasestaa}
\end{equation}
in which the electric susceptibility kernel equals
\begin{equation}
\begin{split}
&\chi(\gamma,t)={N(t)\over\epsilon_0}\alpha(\gamma,t)=\cr
&{N(t)\over\epsilon_0}{e^2\over m\sqrt{\omega_{\rm{n}}^2-{\Gamma_{\rm{D}}^2\over4}}}\exp\big(-{\Gamma_{\rm{D}}\over2}\gamma\big)\sin\big(\sqrt{\omega_{\rm{n}}^2-{\Gamma_{\rm{D}}^2\over4}}\gamma\big).
\end{split}
\label{eqchikkknnnNNN}
\end{equation}
Here, in Eq.~\eqref{eqchikkknnnNNN}, note that we take the polarizability kernel from Eq.~\eqref{appendixfullderi}. Before proceeding, we draw the attention to Eq.~\eqref{eq:pdntptsasestaa} which describes the polarization density. In this equation, we simply wrote $\mathbf{P}(t)$ as the multiplication of the electric dipole moment and the electron density. To do that, firstly, we should assume that by varying the number of electrons (per unit volume) in time, the time-varying material remains homogeneous meaning that the electric susceptibility (or permittivity) does not depend on the position vector $\mathbf{r}$: $\chi(\mathbf{r},\gamma,t)=\chi(\gamma,t)$. Secondly, we should also suppose that the process of changing the electron density in time does not affect the velocities of electrons. Within these assumptions,  Eq.~\eqref{eq:pdntptsasestaa} is valid, and, in fact, we can write $\mathbf{P}(t)=N(t)\mathbf{p}(t)$. Now, by knowing the electric susceptibility kernel from Eq.~\eqref{eqchikkknnnNNN}, we readily find the relative permittivity kernel of the effective medium as 
\begin{equation}
\begin{split}
&\epsilon(\gamma,t)=\delta(\gamma)+\cr
&{N(t)\over\epsilon_0}{e^2\over m\sqrt{\omega_{\rm{n}}^2-{\Gamma_{\rm{D}}^2\over4}}}\exp\big(-{\Gamma_{\rm{D}}\over2}\gamma\big)\sin\big(\sqrt{\omega_{\rm{n}}^2-{\Gamma_{\rm{D}}^2\over4}}\gamma\big),
\end{split}
\label{eqpppkkkktgNN}
\end{equation}
where $\delta(\gamma)$ is the one-dimensional Dirac delta function. Since we have the kernels from the above equations, Eqs.~\eqref{eqchikkknnnNNN} and \eqref{eqpppkkkktgNN}, we can apply the Fourier transform  Eq.~\eqref{eq:susomegatF} and calculate the temporal complex relative permittivity defined in Eqs.~\eqref{eq:DepsEomegat} and \eqref{eq:repersusomegat}. The result reads 
\begin{equation}
\epsilon_{\rm{T}}(\omega,t)=1+{\omega^2_{\rm{p}}(t)\over\omega_{\rm{n}}^2-\omega^2+j\Gamma_{\rm{D}}\omega},
\label{eqqq:tcrp}
\end{equation}
in which 
\begin{equation}
\omega^2_{\rm{p}}(t)={e^2\over\epsilon_0m}N(t)
\end{equation}
is the time-dependent plasma frequency. The expression in Eq.~\eqref{eqqq:tcrp} is complex-valued and explicitly depends on time indicating the nonstationarity characteristic. Substituting $\omega_{\rm n}=0$ (free-electron plasma) we arrive to the conventionally used expression for the effective permittivity of plasma with varying electron density, e.g.~\cite{Ohler1999M}. The only difference with the stationary case is that in the formula the plasma frequency explicitly depends on time.   The reason is due to the low-density approximation that we have made. Within this approximation, the damping coefficient and the natural frequency are considered to be constant in time. Therefore, the polarizability kernel is the same as the one written for the stationary scenario, and consequently the same kind of dispersion is observed. 

Let us consider a more general case when  the damping coefficient and the natural frequency also change in time. In this case, $\epsilon_{\rm{T}}(\omega,t)$ can be dramatically different. Specific dependencies of the effective parameters in \eqref{eqqq:tcrp} are determined by  the plasma structure and can be set as empirical parameters. As a particular example, here we assume that $h(t,\tau)$ is  a product of two functions $K(t)$ and $L(\tau)$ which depend on single independent variables $t$ and $\tau$, respectively. Since $h(t,t)=0$, according to the initial condition in Eq.~\eqref{eq:mtmio}, one also assumes a multiplier in form   $(t-\tau)^n$. Thus, we consider the time-varying dispersion kernel in form 
\begin{equation}
h(t,\tau)=(t-\tau)^nK(t)L(\tau). 
\label{eq:httfullexamplernsm}
\end{equation}
Contemplating the second expression in Eq.~\eqref{eq:mtmio}, we find that \begin{equation}
n(t-\tau)^{n-1}K(\tau)L(\tau)=e^2/m, 
\end{equation}
in which $t$ and $\tau$ must be equal. Due to this feature ($t=\tau$), we can conclude that if $n\neq 1$, the above identity does not hold for nonzero functions of $K(t)$ and $L(\tau)$. Therefore, the only condition for holding the identity occurs when $n$ becomes equal to unity, and, as a result, only under this condition there is a nonzero solution for $h(t,\tau)$. Now, in the case of $n=1$, satisfying the second initial condition determines the function $L(\tau)$ as inversely proportional to $K(\tau)$ such that $L(\tau)={e^2/[mK(\tau)]}$. Using the partial differential equation (the first expression) in Eq.~\eqref{eq:mtmio} and substituting 
\begin{equation}
h(t,\tau)={e^2\over m}(t-\tau){K(t)\over K(\tau)},
\end{equation}
we find the corresponding function $K(t)$: 
\begin{equation}
K(t)=\exp\Big(-\int{\Gamma_{\rm{D}}(t)\over2} dt\Big),
\label{eq:ktgamma}
\end{equation}
with an important constraint:
\begin{equation}
\omega^2_{\rm{n}}(t)={1\over4}\Big[\Gamma^2_{\rm{D}}(t)+2{d\Gamma_{\rm{D}}(t)\over dt}\Big].
\label{eq:wngdrel}
\end{equation}
This equation shows that in this case the temporal variation of the natural frequency fully depends on the temporal variation of the damping coefficient. 

Next, let us assume a free-electron model so that $\omega_{\rm{n}}=0$. Based on Eq.~\eqref{eq:wngdrel}, this assumption forces the damping coefficient to vary homographically as $\Gamma_{\rm{D}}(t)=2\Gamma_0/(1+\Gamma_0t)$, which results in $K(t)=1/(1+\Gamma_0t)$, according to Eq.~\eqref{eq:ktgamma}. With  $h(t,\tau)=(e^2/m)(t-\tau)(1+\Gamma_0\tau)/(1+\Gamma_0t)$, the electric polarizability kernel is given by
\begin{equation}
\alpha(\gamma,t)={e^2\over m}\gamma\Big(1-{\Gamma_0\gamma\over1+\Gamma_0t}\Big).
\label{eq:polartemptVdamping}
\end{equation}
Having the polarizability kernel from Eq.~\eqref{eq:polartemptVdamping}, and after some algebraic manipulations, the relative permittivity kernel is expressed as
\begin{equation}
\epsilon(\gamma,t)=\delta(\gamma)+{N(t)\over\epsilon_0}{e^2\over m}\gamma\Big(1-{\Gamma_0\gamma\over1+\Gamma_0t}\Big).
\label{gen_ker}
\end{equation}
As a sanity check, if $\Gamma_0=0$ and $N(t)$ are  time-invariant, we obtain the  conventional stationary lossless Drude model:
\begin{equation}
\epsilon(\gamma)=\delta(\gamma)+{N\over\epsilon_0}{e^2\over m}\gamma.
\end{equation}

Let us again apply the Fourier transform \eqref{eq:susomegatF} and find the temporal complex relative permittivity which corresponds to kernel \eqref{gen_ker}. In accordance with the properties of Fourier transform, since  
\begin{equation}
\int_0^{+\infty}(-j\gamma)^n\exp(-j\omega\gamma)d\gamma={d^n\over d\omega^n}\Big({1\over j\omega}\Big),
\end{equation}
we find that
\begin{equation}
\begin{split}
&\epsilon_{\rm{T}}(\omega,t)=1-{\omega^2_{\rm{p}}(t)\over\omega^2}-j{2\omega^2_{\rm{p}}(t)\Gamma_0\over\omega^3(1+\Gamma_0t)},\cr
&{\rm{or}}\cr 
&\epsilon_{\rm{T}}(\omega,t)=1-{\omega^2_{\rm{p}}(t)\over\omega^2}-j{\omega^2_{\rm{p}}(t)\over\omega^3}\Gamma_{\rm{D}}(t).
\end{split}
\label{slozhno}
\end{equation}
The temporal permittivity has the imaginary part which is time-dependent, and is negative indicating that the medium is lossy. Comparing the above expression with the conventional stationary Drude model
\begin{equation}
\epsilon_{\rm{Drude}}(\omega)=1-{\omega_{\rm{p}}^2\over\omega^2}{1\over1+({\Gamma_{\rm{D}}\over\omega})^2}-j{\omega_{\rm{p}}^2\over\omega^3}{\Gamma_{\rm{D}}\over1+({\Gamma_{\rm{D}}\over\omega})^2},
\end{equation}
we explicitly observe how the 
effective permittivity of plasma with a time-varying damping coefficient cannot be found by simply assuming that $\Gamma_{\rm{D}}$ depends on time in the conventional Drude formula: Time variations of the damping coefficient $\Gamma_{\rm{D}}(t)$ modifies the real and imaginary parts of the relative permittivity in a different way, as is seen from Eq.~\eqref{slozhno}. 

Finally, before finishing this section, we point out an issue about numerical simulations. The time-invariant (stationary) frequency-dispersive media have been thoroughly studied, and there are well developed time-domain full-wave electromagnetic simulation tools such as Ansys HFSS, CST Microwave Studio, and COMSOL Multiphysics. However, up to our knowledge, there are no numerical tools that could ``properly" simulate dispersive time-varying (nonstationary) media. We hope that this work will help developing such numerical methods which would allow studying electromagnetic processes in dispersive and time-varying media (including time-space modulated metamaterials).

\section{Conclusions}
\label{secconclusion}

We have theoretically studied the fundamental principles related to the electric polarizability of arbitrary dipolar linear particles whose characteristics are varying in time due to some external force. 
Since at every moment of time the time-varying particle is different, the polarizability additionally depends on the observation time (the one at which we measure the dipole moment). Importantly, this time-dependent polarizability is not fully determined by the particle parameters at the observation moment, this polarizability is a causal-response parameter which depends on the whole history of the particle. This is in contrast with a stationary particle whose polarizability depends only on the delay time between the excitation and observation moments.

For time-harmonic excitations, we demonstrated that the instantaneous dipole moment is found as the real part of a complex-valued temporal function that is multiplied by the complex amplitude of the field and the time-harmonic exponential factor. This temporal response function is the Fourier transform of the polarizability kernel with respect to the delay time ($\gamma\rightarrow\omega$). We named this function as \emph{temporal complex polarizability} and explained some of its salient properties. Importantly, using the notion of temporal complex polarizability, we introduced the second Fourier transform that is with respect to the observation time ($t\rightarrow\Omega$). By this way, we could describe the dipole moment completely in the frequency domain. 

Next, we considered a nonstationary particle that is located in free space. By employing the R{\"u}denberg equation (the Hertzian dipole model), we presented a methodical approach to determine the polarizability kernel and the temporal complex polarizability. Having the polarizability, we studied the classical interaction of the dipole with the incident time-harmonic electromagnetic wave. Therefore, in terms of the temporal complex polarizability, we contemplated the instantaneous powers that are extracted by the dipole and scattered from the dipole. 

Finally, we took one step forward and considered the dipole particle as a constituent of a time-varying material. This time, we focused on the classical bound electron model and derived the corresponding equations for finding the polarizability. To do that, we used the equation of motion (or the Drude-Lorentz equation) and assumed that the damping coefficient and the natural frequency are varying in time. Afterwards, we moved towards effective material parameters, and for particular example cases, we derived the effective permittivity of the time-varying medium comprising bound or free electrons. It is observed that this model for describing the effective permittivity is significantly different from the conventional Drude-Lorentz formula with time-dependent parameters.  


\section*{ACKNOWLEDGMENTS}
This work was supported by the Academy of Finland under grant 330260. 
M.S.M.~wishes to acknowledge the support of Ulla Tuominen Foundation. Also, the authors thank V.~Asadchy and A.~Sihvola for their invaluable comments. In addition, M.S.M. thanks X.~Wang for helping to prepare the figure of the paper. 


\appendix

\section{Polarizability of a classical bound electron for the stationary scenario}
Let us present the rigorous derivation of Eq.~\eqref{appendixfullderi}. The first expression in Eq.~\eqref{eq:mtmio} results in 
\begin{equation}
\begin{split}
&{d^2h(u)\over du^2}+\Gamma_{\rm{D}}{dh(u)\over du}+\omega_{\rm{n}}^2h(u)=0\rightarrow h(u)=\exp({-{\Gamma_{\rm{D}}\over2}}u)\cr
&\times \Big(H_1\cos(\sqrt{\omega_{\rm{n}}^2-{\Gamma_{\rm{D}}^2\over4}}u)+H_2\sin(\sqrt{\omega_{\rm{n}}^2-{\Gamma_{\rm{D}}^2\over4}}u)\Big).
\end{split}
\end{equation}
Here, there are two unknown coefficients, $H_1$ and $H_2$, which should be determined from the other two remaining expressions in Eq.~\eqref{eq:mtmio}. In principle, the second and the third expressions, as mentioned, are the initial conditions for determining a specific solution for $h(t,\tau)$. From the second expression, we deduce that
\begin{equation}
{dh(u)\over du}\vert_{u=0}={e^2\over m}\rightarrow-{\Gamma_{\rm{D}}\over2}H_1+\sqrt{\omega_{\rm{n}}^2-{\Gamma_{\rm{D}}^2\over4}}H_2={e^2\over m},
\label{eq:taut2nd}
\end{equation}
and from the last expression in Eq.~\eqref{eq:mtmio}, we conclude that 
\begin{equation}
h(0)=0\rightarrow H_1=0.
\end{equation}
Finally, by combining the above results, $h(t,\tau)$ and the polarizability are given by
\begin{equation}
\begin{split}
&h(t,\tau)=h(t-\tau)=\cr
&{e^2\over m\sqrt{\omega_{\rm{n}}^2-{\Gamma_{\rm{D}}^2\over4}}}\exp\big(-{\Gamma_{\rm{D}}\over2}(t-\tau)\big)\sin\big(\sqrt{\omega_{\rm{n}}^2-{\Gamma_{\rm{D}}^2\over4}}(t-\tau)\big),
\end{split}
\end{equation}
and 
\begin{equation}
\begin{split}
&\alpha(\gamma,t)=h(t,t-\gamma)=\cr
&{e^2\over m\sqrt{\omega_{\rm{n}}^2-{\Gamma_{\rm{D}}^2\over4}}}\exp\big(-{\Gamma_{\rm{D}}\over2}\gamma\big)\sin\big(\sqrt{\omega_{\rm{n}}^2-{\Gamma_{\rm{D}}^2\over4}}\gamma\big),
\end{split}
\end{equation}
respectively.


\section{Inverse polarizability and instantaneous reactive powers} 
\label{appeninvpolinsrpow}
Utilizing the introduced notion of the temporal complex polarizability, we can find the instantaneous scattered power and the total extracted power. Here, we remind an important relation for the imaginary part of the inverse polarizability which is known for stationary dipoles. In the lossless regime and in the time-averaged perspective, by writing that $\overline{S^{\rm{stationary}}}=\overline{S^{\rm{stationary}}_{\rm{rad}}}$, we simply derive the following expression: 
\begin{equation}
\Im\Big[{1\over\alpha_{\rm{T}}}\Big]={k_0^3\over6\pi\epsilon_0},
\label{eq:polimin}
\end{equation}
where $k_0$ and $\epsilon_0$ are the free-space wavenumber and permittivity, respectively. However, under nonstationary conditions, the above equation is not true, and finding a similar relation is not straightforward. This is due to the fact that the conservation of instantaneous power is {\it not} simply $S(t)=S_{\rm{rad}}(t)$ even for the stationary dipole. The stored reactive electric and magnetic energy per unit time should be also taken into account. In other words, by neglecting Ohmic losses, $S(t)=S_{\rm{rad}}(t)+S_{\rm{reactive}}(t)$, where $S_{\rm{reactive}}(t)=S_{\rm{electric}}(t)+S_{\rm{magnetic}}(t)$. To perceive this expression, as an example, let us consider a nonstationary Hertzian dipole loaded with a reactive element (e.g.,~an inductance $L_{\rm{load}}$) changing in time. By using the circuit theory, we can describe the instantaneous reactive powers, $S_{\rm{electric}}(t)$ and $S_{\rm{magnetic}}(t)$, through the effective parameters of the dipole and the temporally modulated reactive load. Those effective parameters, which play an important role in the interaction of the dipole with the incident field, are an effective capacitance $C$ and an effective inductance $L$. As it can be expected, the effective capacitance $C$ corresponds to the stored electric energy near to the dipole, and the stored magnetic energy around the dipole is measured by the effective inductance $L$. Based on this explanation, one can conclude that the total electric energy per unit time is equal to $S_{\rm{electric}}(t)=(Q(t)/C)i(t)$ in which $Q(t)$ is the electric charge and $i(t)$ is the induced electric current in the dipole. However, since we assume a time-varying inductance as the load, the total magnetic energy per unit time consists of contributions from the effective inductance and the load. Consequently, $S_{\rm{magnetic}}(t)=L[di(t)/dt]\cdot i(t)+S_{\rm{load}}(t)$, where $S_{\rm{load}}(t)=[d(L_{\rm{load}}(t)i(t))/dt]\cdot i(t)$. In the above example, we stress that in this derivation the length of the dipole is assumed to be constant in time, and the temporal modulation comes only from the time-varying load connected to the dipole. However, a short dipole can be modulated in time in a number of different ways, for example, the dipole length can vary in time or the shape  of a subwavelength particle such as a lossless sphere can change in time. For each specific case, implications of the above equations and expressions are indeed intriguing and need to be carefully investigated.



\begin{thebibliography}{99} 



\bibitem{Faraday} 
M.~Faraday, 
On a peculiar class of acoustical figures; and on certain forms assumed by groups of particles upon vibrating elastic surfaces, 
Phil.~Trans.~R.~Soc.~Lond.~\textbf{121}, 299 (1831). 


\bibitem{Cullen}
A.~L.~Cullen, 
A travelling-wave parametric amplifier, 
Nature~\textbf{181}, 332 (1958).


\bibitem{Tien}
P.~K.~Tien, 
Parametric amplification and frequency mixing in propagating circuits, 
J.~Appl.~Phys.~\textbf{29}, 1347 (1958).


\bibitem{Morgenthaler}
F.~R.~Morgenthaler, 
Velocity modulation of electromagnetic waves, 
IRE Trans.~Microwave Theory Tech.~\textbf{6}, 167 (1958). 


\bibitem{Kamal}
A.~K.~Kamal, 
A parametric device as a nonreciprocal element, 
Proc.~IRE~\textbf{48}, 1424 (1960). 


\bibitem{Currie}
M.~R.~Currie, and R.~W.~Gould, 
Coupled-cavity traveling-wave parametric amplifiers: Part I-analysis, 
Proc.~IRE~\textbf{48}, 1960 (1960). 


\bibitem{Simon}
J.~C.~Simon, 
Action of a progressive disturbance on a guided electromagnetic wave, 
IRE Trans.~Microwave Theory Tech.~\textbf{8}, 18 (1960).


\bibitem{Edmondson} 
J.~R.~Macdonald, and D.~E.~Edmondson, 
Exact solution of a time-varying capacitance problem, 
Proc.~IRE~\textbf{49}, 453 (1961). 


\bibitem{Oliner} 
A.~Hessel, and A.~A.~Oliner, 
Wave propagation in a medium with a progressive sinusoidal disturbance, 
IRE Trans.~Microwave Theory Tech.~\textbf{9}, 337 (1961).


\bibitem{Anderson}
B.~D.~O.~Anderson, and R.~W.~Newcomb, 
On reciprocity and time-variable networks, 
Proc.~IEEE~\textbf{53}, 1674 (1965). 


\bibitem{Holberg}
D.~E.~Holberg, and K.~S.~Kunz, 
Parametric properties of fields in a slab of time-varying permittivity, 
IEEE Trans.~Antennas Propag.~\textbf{14}, 183 (1966).


\bibitem{TM-isolation1} 
Z.~Yu, and S.~Fan, 
Complete optical isolation created by indirect interband photonic transitions, 
Nature Photonics~\textbf{3}, 91 (2009). 


\bibitem{TM-isolation2}
D.~L.~Sounas, and A.~Al{\`u}, 
Angular-momentum-biased nanorings to realize magnetic-free integrated optical isolation,  
ACS Photonics~\textbf{1}, 198 (2014).


\bibitem{TM-nonreciprocity3} 
Y.~Hadad, D.~L.~Sounas, and A.~Al{\`u}, 
Space-time gradient metasurfaces, 
Phys.~Rev.~B~\textbf{92}, 100304(R) (2015).


\bibitem{TM-nonreciprocity1} 
D.~L.~Sounas, and A.~Al{\`u}, 
Non-reciprocal photonics based on time modulation, 
Nature Photonics~\textbf{11}, 774 (2017). 


\bibitem{TM-nonreciprocity2} 
T.~T.~Koutserimpas, and R.~Fleury, 
Nonreciprocal gain in non-Hermitian time-floquet systems, 
Phys.~Rev.~Letters~\textbf{120}, 087401 (2018). 


\bibitem{TM-nonreciprocity4}
S.~Taravati, N.~Chamanara, and C.~Caloz, 
Nonreciprocal electromagnetic scattering from a periodically space-time modulated slab and application to a quasisonic isolator, 
Phys.~Rev.~B~\textbf{96}, 165144 (2017). 

\bibitem{XuchennnnWWW} 
X.~Wang, G.~Ptitcyn, V.~S.~Asadchy, A.~D{\'i}az-Rubio, M.~S.~Mirmoosa, S.~Fan, and S.~A.~Tretyakov, 
Nonreciprocity in bianisotropic systems with uniform time modulation, 
Phys.~Rev.~Lett.~\textbf{125}, 266102 (2020).


\bibitem{TM-frequencyconvMosallaeiSalary}
M.~M.~Salary, S.~Farazi, and H.~Mosallaei, 
A dynamically modulated all-dielectric metasurface doublet for directional harmonic generation and manipulation in transmission, 
Adv.~Optical~Mater.~\textbf{7}, 1900843 (2019).


\bibitem{TM-frequencytranslationGrbic} 
Z.~Wu, and A.~Grbic, 
Serrodyne frequency translation using time-modulated metasurfaces, 
IEEE Trans.~Antenn.~Propag.~\textbf{68}, 1599 (2020). 


\bibitem{TM-frequency1}
M.~Liu, D.~A.~Powell, Y.~Zarate, and I.~V.~Shadrivov, 
Huygens’ metadevices for parametric waves, 
Phys.~Rev.~X~\textbf{8}, 031077 (2018).


\bibitem{TM-wavefront1}
M.~M.~Salary, S.~Jafar-Zanjani, and H.~Mosallaei, 
Electrically tunable harmonics in time-modulated metasurfaces for wavefront engineering, 
New J.~Phys.~\textbf{20}, 123023 (2018). 


\bibitem{TM-wavefront2}
N.~Chamanara, Y.~Vahabzadeh, and C.~Caloz, 
Simultaneous control of the spatial and temporal spectra of light with space-time varying metasurfaces, 
IEEE Trans.~Anten.~Propag.~\textbf{67}, 2430 (2019). 


\bibitem{TM-beam}
S.~Taravati, and A.~A.~Kishk, 
Dynamic modulation yields one-way beam splitting, 
Phys.~Rev.~B~\textbf{99}, 075101 (2019).

\bibitem{TM-Energy}
M.~S.~Mirmoosa, G.~A.~Ptitcyn, V.~S.~Asadchy, and S.~A.~Tretyakov, 
Time-varying reactive elements for extreme accumulation of electromagnetic energy, 
Phys.~Rev.~Applied~\textbf{11}, 014024 (2019).



\bibitem{Fleury222}
T.~T.~Koutserimpas, A.~Al{\`u}, and R.~Fleury, 
Parametric amplification and bidirectional invisibility in PT-symmetric time-Floquet systems, 
Phys.~Rev.~A~\textbf{97}, 013839 (2018). 


\bibitem{Fleury333}
T.~T.~Koutserimpas, and R.~Fleury, 
Electromagnetic waves in a time periodic medium with step-varying refractive index, 
IEEE Trans.~Anten.~Propag.~\textbf{66}, 5300 (2018).


\bibitem{TM-Matching}
A.~Shlivinski, and Y.~Hadad, 
Beyond the Bode-Fano bound: Wideband impedance matching for short pulses using temporal switching of transmission-line parameters, 
Phys.~Rev.~Letters~\textbf{121}, 204301 (2018).


\bibitem{TM-Radiation1}
G.~Ptitcyn, M.~S.~Mirmoosa, and S.~A.~Tretyakov, 
Time-modulated meta-atoms, 
Phys.~Rev.~Research~\textbf{1}, 023014 (2019). 

\bibitem{TM-Radiation2}
M.~S.~Mirmoosa, G.~A.~Ptitcyn, R.~Fleury, and S.~A.~Tretyakov, 
Instantaneous radiation from time-varying electric and magnetic dipoles, 
Phys.~Rev.~A~\textbf{102}, 013503 (2020). 

\bibitem{polarizability-tensors} 
A.~Serdyukov, I.~Semchenko, S.~Tretyakov, and A.~Sihvola, 
{\it Electromagnetics of Bi-anisotropic Materials: Theory and Applications} (Gordon and Breach Science Publishers, Amsterdam, 2001). 

\bibitem{CalozTVMM1}
C.~Caloz, and Z.-L.~Deck-L{\'e}ger, 
Spacetime metamaterials--Part I: General concepts, 
IEEE Trans.~Antenn.~Propag.~\textbf{68}, 1569 (2020).


\bibitem{CalozTVMM2}
C.~Caloz, and Z.-L.~Deck-L{\'e}ger, 
Spacetime metamaterials--Part II: Theory and applications, 
IEEE Trans.~Antenn.~Propag.~\textbf{68}, 1583 (2020). 

\bibitem{SihvolaMF} 
A.~Sihvola, 
{\it Electromagnetic Mixing Formulas and Applications} (The Institution of Electrical Engineers, London, UK, 1999). 

\bibitem{Enghetanano}
V.~Pacheco-Pe{\~n}a, and N.~Engheta, 
Effective medium concept in temporal metamaterials, 
Nanophotonics~\textbf{9}, 379 (2020).


\bibitem{communicationengLTVS} 
H.~D'Angelo, 
{\it Linear Time-Varying Systems: Analysis and Synthesis} (Allyn \& Bacon, Boston, 1970). 


\bibitem{ZADEHAIP} 
L.~A.~Zadeh, 
Frequency analysis of variable networks, 
Proc.~IRE~\textbf{38}, 291 (1950).



\bibitem{Landauref} 
L.~D.~Landau, and E.~M.~Lifshitz, 
{\it Electrodynamics of Continuous Media} (Pergamon, Oxford, 1960). 


\bibitem{TEXTBOOKSLTIS} 
J.~Schwinger, L.~L.~DeRaad, Jr., K.~A.~Milton, and W.-Y.~Tsai, 
{\it Classical Electrodynamics} (Perseus, Reading, MA, 1998). 


\bibitem{jackson}
J.~D.~Jackson, 
{\it Classical Electrodynamics} (John Wiley \& Sons, New York, 1999). 

\bibitem{Halevieutv2009}
J.~R.~Zurita-S{\'a}nchez, P.~Halevi, and J.~C.~Cervantes-Gonz{\'a}lez, 
Reflection and transmission of a wave incident on a slab with a time-periodic
dielectric function $\epsilon(t)$, 
Phys.~Rev.~A~\textbf{79}, 053821 (2009). 

\bibitem{Budko2009}
N.~V.~Budko, 
Electromagnetic radiation in a time-varying background medium, 
Phys.~Rev.~A~\textbf{80}, 053817 (2009).

\bibitem{Halevieutv2016}
J.~S.~Mart{\'i}nez-Romero, O.~M.~Becerra-Fuentes, and P.~Halevi, 
Temporal photonic crystals with modulations of both permittivity and permeability, 
Phys.~Rev.~A~\textbf{93}, 063813 (2016). 

\bibitem{Stepanov1976}
N.~S.~Stepanov, 
Dielectric constant of unsteady plasma, 
Izv.~Vyssh.~Uchebn.~Zaved.,~Radiofiz.~\textbf{19}, 960 (1976). 


\bibitem{rudenberg1907} 
R.~R{\"u}denberg, 
Der Empfang elektrischer Wellen in der drahtlosen Telegraphie, 
Annalen der Physik~\textbf{330}, 446 (1908). 

\bibitem{NH2012} 
L.~Novotny, and B.~Hecht, 
{\it Principles of Nano-Optics} (Cambridge University Press, Cambridge, UK, 2012). 

\bibitem{Tretyakovplasmonics} 
S.~Tretyakov,
Maximizing absorption and scattering by dipole particles,
Plasmonics~\textbf{9}, 935 (2014).

\bibitem{Ohler1999M}
S.~G.~Ohler, B.~E.~Gilchrist, and A.~D.~Gallimore, 
Electromagnetic signal modification in a localized high-speed plasma flow: Simulations and experimental validation of a stationary plasma thruster, 
IEEE Transactions on Plasma Science~\textbf{27}, 587 (1999).



\end{thebibliography}
\end{document}